\documentclass{article}

\usepackage{arxiv}

\usepackage[utf8]{inputenc} 
\usepackage[T1]{fontenc}    
\usepackage{url}            
\usepackage{booktabs}       
\usepackage{amsfonts}       
\usepackage{nicefrac}       
\usepackage{microtype}      
\usepackage{lipsum}
\usepackage{graphicx}
\graphicspath{ {./images/} }
\usepackage{graphicx}
\usepackage{amsmath}
\usepackage{amssymb}
\usepackage{natbib} 

\usepackage{newtxtext}
\usepackage{newtxmath}
\usepackage{bm}
\usepackage{marvosym}

\usepackage[
    colorlinks=true, 
    linkcolor=black, 
    citecolor=black  
]{hyperref}

\title{Extreme vortex gust encounters by a square wing}

\author{
    Hiroto Odaka\thanks{Email for correspondence: hodaka@g.ucla.edu}, \  Luke Smith, \ and Kunihiko Taira \\ \\
    Department of Mechanical and Aerospace Engineering, University of California, Los Angeles, CA 90095, USA
}

\begin{document}
\maketitle
\begin{abstract}
Extreme gust encounters by finite wings with disturbance velocity exceeding their cruise speed remain largely unexplored, while particularly relevant to miniature-scale aircraft. 
This study considers extreme aerodynamic flows around a square wing and the large, unsteady forces that result from gust encounters.
We analyse the evolution of three-dimensional, large-scale vortical structures and their complex interactions with the wing by performing direct numerical simulations at a chord-based Reynolds number of 600.
We find that a strong incoming positive gust vortex induces a prominent leading-edge vortex~(LEV) on the upper surface of the wing, accompanied by tip vortices~(TiVs) strengthened through the interaction. 
Conversely, a strong negative gust vortex induces an LEV on the lower surface of the wing and causes a reversal in TiV orientation.
In both extreme vortex gust encounters, the wing experiences significant lift fluctuations.
Furthermore, we identify two opposing effects of the TiVs on the large lift fluctuations.
First, the enhanced or reversed TiVs contribute to significant lift surges or drops by generating large low-pressure cores near the wing.
Second, the TiVs play a part in attenuating lift fluctuations through enhanced downwash or upwash, formation of an arch vortex, and distortion of vortical structure around the wing corners.
The second effect outweighs the first, resulting in smaller transient lift changes on the finite wing compared to the 2D wing.
We also show that flying above a positive gust vortex or flying below a negative one can mitigate lift fluctuations during encounters.
The current findings provide potential guidance on how TiV dynamics and wing positions could be leveraged to alleviate large transient lift fluctuations experienced by finite wings in severe gust conditions.
\end{abstract}


\section{Introduction}
\label{sec:intro}

Strong gust encounters present a substantial challenge in aviation because of their time-variant, transient dynamics with strong nonlinearity.
Such gust encounters exert large unsteady aerodynamic loads on wings, negatively impacting flight stability and aircraft structures~\citep{fuller1995evolution,wu2019gust,anya_review}.
Especially for small-scale air vehicles, such as drones, the relative scale of gust disturbances can be disproportionally large, a consequence of the small size and low cruise velocity associated with these vehicles~\citep{mueller2003aerodynamics,anya_review}.
For instance, during low-speed maneuver operations of small-scale aircraft, such as takeoff or landing in confined urban environments, the relative strength of gusts can be significantly high.
Typically, when the gust ratio~$G$---the ratio between characteristic gust velocity and cruise velocity---exceeds~1, the aerodynamic environment is considered unflyable.
When $G>1$, the dynamics of the flow is referred to as \textit{extreme aerodynamics}~\citep{fukami2023grasping,taira2025extreme}.
To enable the use of small-scale aircraft for transportation, agriculture, and search-and-rescue~\citep{intro_applicatoin1, intro_applicatoin2,ahirwar2019application,intro_applicatoin3} during adverse weather, it is crucial to advance our understanding of extreme aerodynamics.

To address this issue, past literature studying unsteady aerodynamics associated with gust-wing interactions provides valuable insights.
Early efforts examined the aerodynamics of gust-wing interaction using linear thin airfoil theory~\citep{press1954study,horlock1968fluctuating,goldstein1976complete}.
They treated gusts as weak perturbations, assuming attached flows.
For instance, \cite{atassi1984sears} derived lift formulas for a thin airfoil with small camber at a low angle of attack subjected to a vertical gust, incorporating the effect of the distortion of the gust structure.
Their theoretical analysis was later shown to be in good agreement with experimental results~\citep{cordes2017note}.

More recent studies have explored gust ratios in the range of~$0 < G\lesssim 1$, where massive flow separation occurs around wings.
These studies analysed vortex interactions and aerodynamic force responses for various types of gust encounters~\citep{jones2021overview}, including transverse gusts~\citep{wang2024airfoil}, streamwise gusts~\citep{ma2021unsteady}, and vortex gusts~\citep{martinez2020analysis}. 
For instance, \citet{peng2017asymmetric} experimentally examined load distributions on a 2D wing during vortex gust encounters in the context of blade-vortex interactions.
Furthermore, \citet{herrmann2022gust} designed a closed-loop controller using an actuated trailing-edge flap on an airfoil to attenuate lift fluctuation during vortical gust encounters.
Building upon this body of work, extreme vortex-gust-wing interactions with~$G\gtrsim1$ were recently investigated at a Reynolds number of 100 by~\citet{fukami2023grasping} and~\citet{fukami2024data}. 
They observed massive flow separation with high levels of transient lift change and developed a lift attenuation strategy based on a low-order manifold around which the extreme dynamics evolve.

While the previously mentioned studies mainly focus on two-dimensional or spanwise periodic wings in the absence of tip vortices~(TiVs), there exists some work that examined gust encounters by a finite wing.
\citet{barnes2018clockwise,barnes2018counterclockwise} investigated the evolution of boundary layers and the force load on a finite wing during vortex-gust encounters.
Control strategies, such as passive twist wingtips, for force load attenuation on a finite wing during vertical gust encounters were also proposed by \citet{guo2015gust} and~\citet{he2021passive}.
However, their work is primarily centered around $G\lesssim 1$ at a low angle of attack.
Therefore, building a foundational understanding of gust-finite-wing interactions in a regime of~$G\gtrsim1$ is critical to further advancing the field of extreme aerodynamics.

This study aims to elucidate the aerodynamics of extreme finite-wing-vortex-gust interactions at a Reynolds number of~$600$.
The Reynolds number is sufficiently high to trigger massive flow separation, yet still low enough to preserve a laminar flow.
Note that the core vortical structures observed in massively separated flows maintain topological similarities between low- and high-Reynolds-number regimes~\citep{hunt1978kinematical,dallmann1988three,delery2001robert}, while the varying scales of vortical structures present at high Reynolds numbers significantly complicate flow analysis, especially without an understanding of large-scale vortex behavior.
Furthermore, for extreme aerodynamics, there is a disparity between the advective time scale and the viscous time scale, with the former being considerably smaller than the latter~\citep{taira2025extreme}.
As such, the vortex core dynamics are similar across a range of Reynolds numbers.
By exploring the large-scale vortical dynamics in the absence of turbulence, this work aims to establish a foundation for understanding extreme gust encounters by finite wings, setting a stepping stone for future investigations at higher Reynolds numbers.

In the present study, we perform direct numerical simulations of extreme vortex gust encounters by a square wing and analyse the complex interplay between flow structures, including a gust vortex, wing-surface separation, and TiVs.
Moreover, we identify vortical structures that serve prominent roles in aerodynamics.
The rest of the paper is structured as follows. 
The computational set-up for an extreme vortex-gust encounter by the square wing is described in section~\ref{sec:problemSetup}. 
We present the vortex dynamics of the encounters and identify primary vortical structures responsible for the high lift changes in section~\ref{sec:result}.
We also discuss the attenuation of lift variations due to TiV dynamics and wing positions against an impacting gust vortex, providing insights into mitigation strategies of the transient lift fluctuations.
Concluding remarks are offered in section~\ref{sec:conclusion}.

\section{Problem setup}
\label{sec:problemSetup}

\begin{figure}
\begin{center}
   \includegraphics[width=1\textwidth, scale=1.0]{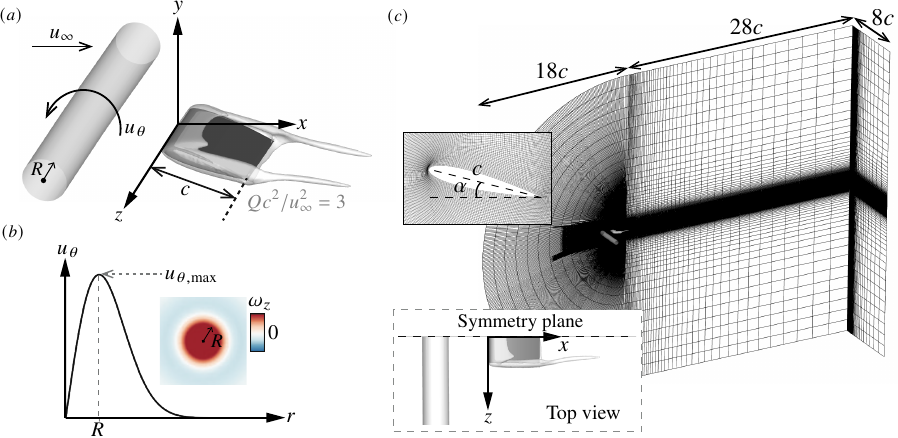}
   \caption{
   $(a)$~NACA0015 square wing encountering a gust vortex modeled as a Taylor vortex.
   Q-criterion isosurface is shown.
   $(b)$~ Circumferential velocity profile and spanwise vorticity distribution of a Taylor vortex with a positive orientation.
   $(c)$~Computational domain and discretization.
   }
\label{fig:fig1}
\end{center}
\end{figure}

We analyse flows around a NACA0015 finite wing impacted by a vortex gust, as presented in figure~\ref{fig:fig1}$(a)$.
The vortex gust is modeled as a spanwise-oriented vortex column with the circumferential velocity profile of a Taylor vortex~\citep{Taylor_vortex}
\begin{equation}
    u_{\theta}=u_{\theta, \text{max}} \frac{r}{R} \exp \left[ \frac{1}{2} \left( 1- \frac{r^2}{R^2} \right)  \right],
\end{equation}
where $u_{\theta, \text{max}}$ is the maximum rotational velocity at radius $r=R$.
This vortex gust is characterized by the gust ratio
\begin{equation}
    G \equiv u_{\theta,\text{max}}/u_{\infty},
\end{equation}
where $u_{\infty}$ is the free-stream velocity.
The circumferential velocity profile and the spanwise vorticity distribution are provided in figure~\ref{fig:fig1}$(b)$.

We consider a square wing~(semi-aspect ratio~$sAR=0.5$) with a straight-cut wingtip at an angle of attack of~$\alpha=14^\circ$ and a chord-based Reynolds number of $Re=u_{\infty}c/\nu=600$, where $c$ is the chord length and $\nu$ is the kinematic viscosity.
The aspect ratio is selected to effectively study tip effects, while the angle of attack is chosen to ensure a leading-edge flow separation is present for the undisturbed flow.
We selected the Reynolds number to examine large-scale vortical dynamics in the absence of turbulence.
The steady baseline~(undisturbed) flow imposes a lift of $C_{L,\text{base}}=0.22$, where the lift coefficient is defined as~$C_L \equiv F_L/(\frac{1}{2} \rho u_\infty^2 bc)$.
Here, $F_L$ is the $y$-component of the force load on the wing, $\rho$ is density, and $b$ is the half-span length.

We perform direct numerical simulations with a compressible flow solver CharLES~\citep{charles2}, a finite-volume solver with second-order and third-order accuracy in space and time, respectively.
The Mach number~$u_\infty/a_\infty=0.1$ is used to minimize compressible flow effects.
The computational domain is shown in figure~\ref{fig:fig1}$(c)$, where we respectively define $x$, $y$, and $z$ as the streamwise, vertical, and spanwise coordinates with the leading edge of the wing root at the origin~$(0,0,0)$.
We impose a Dirichlet boundary condition of $u_\infty/a_\infty=0.1$ at the inlet and farfield boundaries, an adiabatic wall boundary condition at the wing surface, and a sponge boundary condition at the outlet boundary.
Along the $z/c=0$ plane~(wing root), we prescribe a symmetry boundary condition.
The time step is selected such that the local Courant number is less than~1 throughout the entire computational domain.
The present simulation is verified through a grid convergence study and validated with the lift coefficient for similar $sAR=0.5$ and $sAR=2$ wings reported in~\citet{zhang_rapid} and~\citet{ribeiro2023laminar}, as detailed in appendix~\ref{sec:appendix:b}.

We select four values of gust ratios $G= \{-1.2,-3,1.2,3 \}$ with a fixed gust radius of $R/c=0.25$.
We introduce the gust vortex at a fixed streamwise location of $x_0/c=-3$ and varied vertical positions of $y_0/c=\{-0.25,0,0.25\}$ in the fully-converged undisturbed flow and allow it to convect with the freestream.
We define a reference time $\tau=u_\infty t / c$, such that $\tau=0$ corresponds to the time at which the centre of the vortex column would reach the leading edge if there were no diffusion or interaction between the gust and any other flow structures.
For comparison, we also simulate extreme vortex gust encounters with a NACA0015 2D wing over a strictly two-dimensional domain, where we use the same freestream and vortex conditions as in the square wing case.

\section{Results}
\label{sec:result}

Let us begin by examining the lift response experienced by the wings encountering an extreme gust vortex initially introduced at~$(x_0/c,y_0/c)=(-3,0)$. 
The time traces of lift coefficients for $G= \{-1.2,-3,1.2,3 \}$ are shown in figure~\ref{fig:fig2}, where dashed and solid lines represent the 2D and square wings, respectively.
The baseline flow around the 2D wing is unsteady with $\bar{C}_{L,\text{base}}=0.49$, while the baseline flow around the square wing is steady with $C_{L,\text{base}}=0.22$.
Note that the unsteady unperturbed flow for the 2D wing can capture the unsteady flow physics even from the start of the gust encounter.
Although the baseline flow for the 2D wing exhibits unsteady wake, we confirm that the timing of the gust encounter with respect to the oscillation in the baseline lift causes no significant difference in the transient lift during the extreme encounters, similar to a recent study with an infinite-span wing~\citep{fukami2025extreme}.
This is because the large lift fluctuation due to the extreme encounters is much more significant than the lift fluctuation of the undisturbed case.

In the 2D case, the maximum lift load in the vortex encounters exceeds five times the baseline value for $G=1.2$ and eleven times for $G=3$, as shown in figure~\ref{fig:fig2}.
Compared to the 2D wing, the square wing experiences notably lower lift spikes.
However, despite this reduction, the finite wing still undergoes substantial lift fluctuations.
In the following sections, we explain what vortex dynamics are responsible for the large lift changes and what are the key three-dimensional dynamics that cause the reduced lift response compared to the 2D wing.

\subsection{Vortex encounters by the 2D wing}
\label{subsec:3.1}
\begin{figure}
\begin{center}
   \includegraphics[width=0.95\textwidth, scale=1]{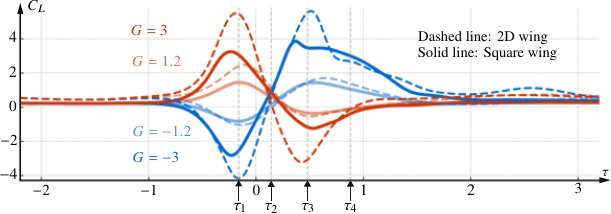}
   \caption{
   Lift history for the 2D~(dashed line) and square~(solid line) wings with $G= \{-1.2,-3,1.2,3 \}$.
   Representative temporal instances are indicated with $\tau=\tau_1$, $\tau_2$, $\tau_3$, and $\tau_4$.
   }
\label{fig:fig2}
\end{center}
\end{figure}

\begin{figure}
\begin{center}
   \includegraphics[width=1\textwidth, scale=1.0]{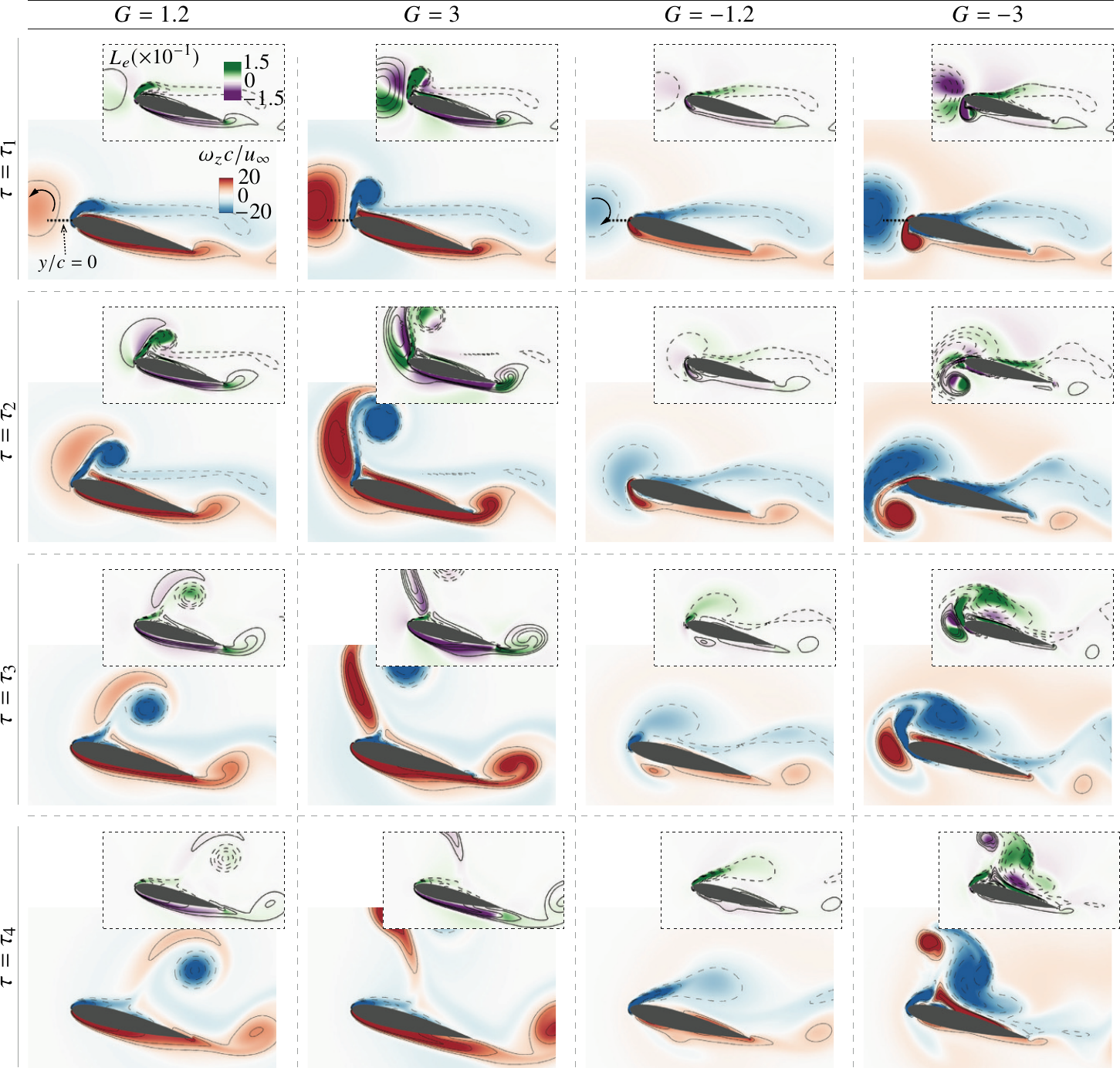}
   \caption{
   Snapshots of spanwise vorticity~$\omega_z$ for the 2D wing with $G= \{-1.2,-3,1.2,3 \}$ at four temporal instances $\tau=\tau_1$ through $\tau_4$ noted in figure~\ref{fig:fig2}.
   Lift elements~$L_e$~(green and purple contours) with lined contours of $\omega_z$ are inserted at the top right of each subplot.
   }
\label{fig:fig3}
\end{center}
\end{figure}

We analyse the flow dynamics during the extreme vortex gust encounters by the 2D wing with $y_0/c=0$.
The evolutions of spanwise vorticity~$\omega_z$ are shown in figure~\ref{fig:fig3} at the four temporal instances from~$\tau=\tau_1$ to~$\tau_4$ noted in figure~\ref{fig:fig2}.
For positive vortex cases~($G=1.2$ and~$3$ in figure~\ref{fig:fig3}), the increasing effective angle of attack due to the incoming gust vortex promotes vorticity generation from the leading edge, forming a large leading-edge vortex~(LEV) above the wing.  
Furthermore, the approaching gust vortex increases the wall-normal velocity gradient on the bottom surface of the wing, increasing the vorticity contained within the boundary layer.
The attached lower boundary layer with an increased level of vorticity flux on the bottom surface, coupled with a separated upper boundary layer, leads to the formation of a trailing-edge vortex~(TEV).
Accordingly, the lift peak rises to $2.47$ and $5.49$ for $G=1.2$ and $G=3$, respectively.

Focusing our attention on times~$\tau=\tau_2$ and~$\tau_3$ for $G=1.2$ and~$3$, the gust vortex is significantly deformed when it impinges upon the leading edge of the wing.
The lower part of the gust vortex merges with the boundary layer on the bottom surface of the wing, while the upper part of the gust vortex convects upward and downstream, forming a vortex-pair-like structure including the LEV.
By $\tau=\tau_3$, the LEV separates from the wing, causing the wing to lose the major lift-generating structures while the negative-lift-contributing structures on the lower surface retain ---the lift coefficient peak drops to $-3.22$ for $G=3$.
By~$\tau=\tau_4$, the leading-edge shear layer gradually starts reforming, restoring the baseline flow state.

The evolutions of 2D, negative vortex cases are similar to that of the positive vortex cases, but with a reversal in the direction of the gust-induced velocity. 
As shown for $G=-1.2$ at~$\tau=\tau_1$ in figure~\ref{fig:fig3}, vertically downward velocity from the approaching gust vortex delays separation above the wing while vorticity generation is promoted below the wing although it does not grow enough to form an LEV.
For $G=-3$, the downward velocity by the approaching gust vortex is so strong that an LEV develops beneath the wing, causing the lift coefficient~$C_L$ to drop below~$-4$.

For both $G=-1.2$ and $-3$, the gust vortex merges with the leading-edge shear layer on the upper surface around $\tau=\tau_2$.
Moreover, for $G=-3$ , the LEV forming below the wing induces counter-rotating vorticity between itself and the wing, generating another vortical structure, as observed at~$\tau=\tau_2$ in figure~\ref{fig:fig3}. 
This structure grows along with the LEV until it disrupts the vortical sheet feeding the LEV, resembling the dynamics observed around a revolving wing~\citep{garmann2014dynamics}.
Afterward, this opposite-signed vortical structure rolls up over the leading edge, eventually forming a secondary LEV on the top side of the wing, as seen at~$\tau=\tau_3$ and $\tau_4$.

To quantify the contribution of vortical structures to lift, we use force element analysis~\citep{chang1992potential,menon2021initiation}, a method to identify flow structures responsible for load generation on the wing.
We derive an auxiliary potential~$\phi_y$ under a boundary condition of $-\bm{n}\cdot\nabla \phi_y=\bm{n}\cdot\bm{e}_y$ on the wing surface, where $\bm{e_y}$ is the unit vector in the $y$ direction.
By integrating the inner product of the Navier–Stokes equations with $\nabla \phi_y$ over the fluid domain, we can express the lift force as
\begin{equation}
    F_L=\int_V\bm{\omega}\times\bm{u}\cdot\nabla\phi_y dV + \frac{1}{Re}\int_S \bm{\omega}\times\bm{n}\cdot \left( \nabla\phi_y+\bm{e}_y \right)dS.
\end{equation}
The first integrand represents volume lift elements while the second integrand represents surface lift elements.
For the current flows at $Re = 600$, the volume lift elements have dominant contributions to lift force, compared to the surface elements~\citep{ribeiro2023laminar}.
We hereafter refer to the volume lift element as lift element~$L_e$.

The lift element plot is inserted at the top right of each subplot in figure~\ref{fig:fig3}, where green and purple contours correspond to lift-increasing and lift-decreasing elements, respectively. 
For the positive vortex cases, the LEV is the prominent contributor to positive lift, while the boundary layer on the lower surface and the gust vortex merged with it primarily yield negative lift.
For instance, the lift contribution from the LEV at~$\tau=\tau_1$ for $G=3$ totals~$4.63$, where the lift elements with $L_e\geqslant 0.015$ inside the vortex are integrated and scaled by $\frac{1}{2} \rho u_\infty^2 c$.
In contrast, for the negative gust vortex cases, the LEV and shear layer below the wing mainly contribute to negative lift, whereas the gust vortex merged with the shear layer on the top surface and the secondary LEV predominantly generate positive lift;
e.g., at~$\tau=\tau_1$ for $G=-3$, the LEV below the wing accounts for a total negative lift contribution of~$-2.63$~(integration of the lift elements with $L_e\leqslant -0.015$ inside the vortex).
In the next subsection, we explore the effects of the wing being finite (in the spanwise direction) on the vortex dynamics and lift fluctuations in extreme vortex gust encounters.

\begin{figure}
\begin{center}
   \includegraphics[width=0.9\textwidth, scale=1]{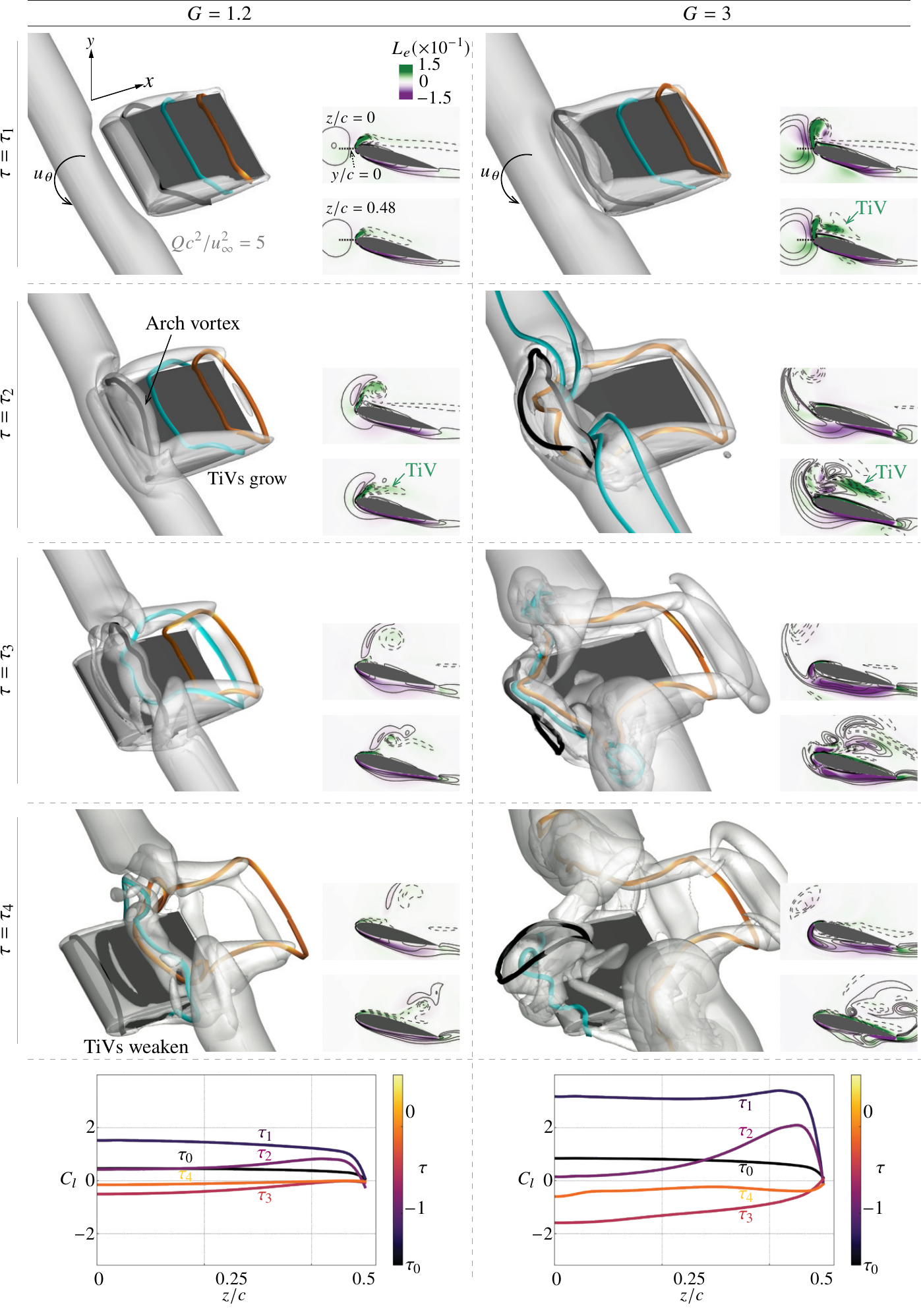}
   \caption{
   Top-port views for the square wing cases with $G=1.2$ and~$3$ at the four temporal instances noted in figure~\ref{fig:fig2}.
   Q-criterion iso-surface is shown in gray with three representative vortex lines coloured in black, aqua, and orange.
   Spanwise slices of lift elements~$L_e$~(colour contours) with $\omega_z$~(line contours) along the root~$z/c=0$ and near the tip~$z/c=0.48$ are shown on the right of each subplot.
   Sectional lift distributions at~$\tau_0=-0.6$ and~$\tau=\tau_1$ through $\tau_4$ are presented at the bottom.
   }
\label{fig:fig4}
\end{center}
\end{figure}

\subsection{Positive vortex encounters by the square wing}
\label{subsec:3.2}
Let us examine the 3D vortex gust encounters by the square wing, focusing on vortical structures responsible for large lift variation and its attenuation compared to the 2D wing.
The temporal evolution of vortical structures for the positive gust vortex cases with~$y_0/c=0$ is presented in figure~\ref{fig:fig4}, where we visualize a Q-criterion isosurface in gray and three representative vortex lines in black, aqua, and orange at each instance.
Note that the plotted vortex lines are shown to facilitate describing the details of vortical structures, and lines in the same colour do not necessarily represent the same structures being tracked over time.
Each panel also presents spanwise slices of lift element~$L_e$~(coloured contour) and $\omega_z$~(lined contour) along the root~$z/c=0$ and near the wingtip~$z/c=0.48$ on the right.

We begin by discussing the effects of TiVs, one of the primary vortices present in flows around the finite wing but absent in the 2D wing cases.
First, the positive gust vortex lowers its vertical position, likely due to the exposure to the downward velocity induced by the TiVs until the impact.
This effect can be seen for both $G=1.2$ and $3$ cases in gust position relative to $y/c=0$ at~$\tau=\tau_1$ in figure~\ref{fig:fig3} and the spanwise slices in figure~\ref{fig:fig4}.
The lowered impact position weakens LEV development, reducing the lift contribution of the LEV compared to the 2D wing.

When the gust vortex is approaching the wing, the TiVs grow in size and strength, particularly toward the leading edge.
This occurs because the enhanced upward velocity from the incoming gust vortex increases the pressure difference between the upper and lower surfaces of the wing.
The strengthened TiVs locally produce enhanced downwash, as indicated in a larger sectional lift drop near the tip at~$\tau=\tau_1$ than~$\tau=\tau_0$ in the last row of figure~\ref{fig:fig4}, which displays the time evolution of sectional lift distribution at~$\tau_0=-0.6$ and~$\tau=\tau_1$ to $\tau_4$. 
An additional discussion on the enhanced downwash is provided in appendix~\ref{sec:appendix:a}.
After the gust vortex passes the wing, its vertical velocity around the wing directs downward.
This reduces the pressure difference between both sides of the wing, weakening the TiVs.
The TiV evolution can be seen in figure~\ref{fig:fig4}, where the TiVs strengthen as the gust vortex approaches the wing and weaken after it moves past the wing.

Next, let us examine how the vortices form and interact with each other around the finite wing.
As observed at $\tau_1$ in figure~\ref{fig:fig4}, the LEV, TiVs, and boundary layer on the wing bottom form a closed loop of a vortex line.
By~$\tau=\tau_2$, the LEV around the tips is anchored to the wing corners, transforming into an arch vortex.
For $G=1.2$ at~$\tau=\tau_3$, arch vortices convect above the wing, with their legs connecting to the TiVs~(the vortex line in aqua) or the boundary layer on the bottom surface~(the vortex line in black).
Over time, the arch vortices convect downstream, exhibiting kinks near the tip~($\tau=\tau_4$).

For $G=3$ at $\tau=\tau_2$, part of the LEV becomes connected with the upper portion of the gust vortex, forming a vortex loop~(the vortex line in black).
Another part of the LEV forms a loop with the TEV~(the vortex line in orange).
The aqua-coloured vortex line depicts part of the gust vortex connecting to vortical structures that have opposite-sign spanwise vorticity outside of the gust vortex core.
At $\tau=\tau_3$, the LEV connects to the gust vortex~(black), the boundary layer below the wing~(aqua), and the TEV~(orange).
By $\tau=\tau_4$, the vortical structure represented by the vortex line in black distorts around the arch vortex.

Let us turn our attention to the vortical structures responsible for the large lift change through the lens of lift elements in figure~\ref{fig:fig4}.
We find similarities with the 2D wing cases.
The LEV remains the primary contributor to positive lift, while the boundary layer on the bottom surface and the gust vortex merged with it play a dominant role in negative lift;
e.g., the lift elements of the LEV totals~1.91 at~$\tau=\tau_1$, where the volume integral of lift elements with $L_e\geqslant 0.015$ inside the vortex is performed and normalized by $\frac{1}{2} \rho u_\infty^2 bc$.
However, in the finite wing case, another prominent source of positive lift emerges: 
the TiVs, particularly notable at~$\tau=\tau_1$ and~$\tau_2$.
While the TiVs cause sectional lift reduction due to their local downwash, the TiVs, strongly amplified by the impacting gust vortex, generate large low-pressure cores above the wing, enhancing their local contribution to vortex lift~\citep{lee2012vorticity,smith2024effect}.
Specifically, at~$z/c=0.48$, the sectional lift attributed to the TiV for~$G=3$ is~0.89, which is nearly three-quarters of that by the LEV~($=1.24$), where we integrate lift elements with $L_e\geqslant 0.015$ inside each vortex along the spanwise slice.
Although the lift contribution of the TiVs is pronounced only around the tip, their contribution to the total lift is~0.34 at~$\tau=\tau_1$, which is over 10\% of it.
The positive lift effect of the TiVs is also observed in the sectional lift distribution visualized in figure~\ref{fig:fig4}, where, with the growth of the TiVs, the section lift has a positive peak near the wingtips, a trend particularly pronounced at~$\tau=\tau_2$.

We further examine key three-dimensional vortex dynamics that attenuate lift fluctuation compared to the 2D case.
First, the effective angle of attack is reduced likely due to the enhanced downwash by the strengthened TiVs, decreasing positive lift around the first peak at~$\tau=\tau_1$.
Additionally, the lower impact position of the gust vortex and the formation of the arch vortex suppress LEV development~\citep{lee2012vorticity}, further reducing positive lift.
The arch vortex also plays a role in attenuating the negative peak at~$\tau=\tau_3$ by keeping the LEV closer to the upper surface near the tips.
Moreover, spanwise vorticity generation in the lower-surface boundary layer is suppressed around the wingtips while the gust vortex merging with the boundary layer distorts around the wing corners, leading to a decreased level of spanwise vorticity on the underside of the wing---and consequently, a reduction in negative lift elements compared to the 2D wing.
This loss of negative lift elements near the tips leads to an increasing trend in the sectional lift toward the wingtip, as seen particularly at~$\tau=\tau_3$ in the last row of figure~\ref{fig:fig4}.

We therefore find two opposing effects of the TiVs on the large lift fluctuations induced by an extreme positive gust vortex.
First, they contribute to positive lift by creating large low-pressure cores near the tips above the wing.
Second, they attenuate the large lift surge and drop, likely due to lowering the impact position of an incoming gust vortex, enhancing downwash with their growth, and generating arch vortices.
The second effect, combined with reduced spanwise vorticity generation in the lower-surface boundary layer and the distortion of the gust vortex around the tips, dominates the first, resulting in the finite wing undergoing smaller lift fluctuations than the 2D wing.
This suggests that further promoting TiVs during interaction with a positive gust vortex could lead to a greater attenuation of large transient lift fluctuations.

\begin{figure}
\begin{center}
   \includegraphics[width=1\textwidth, scale=1.0]{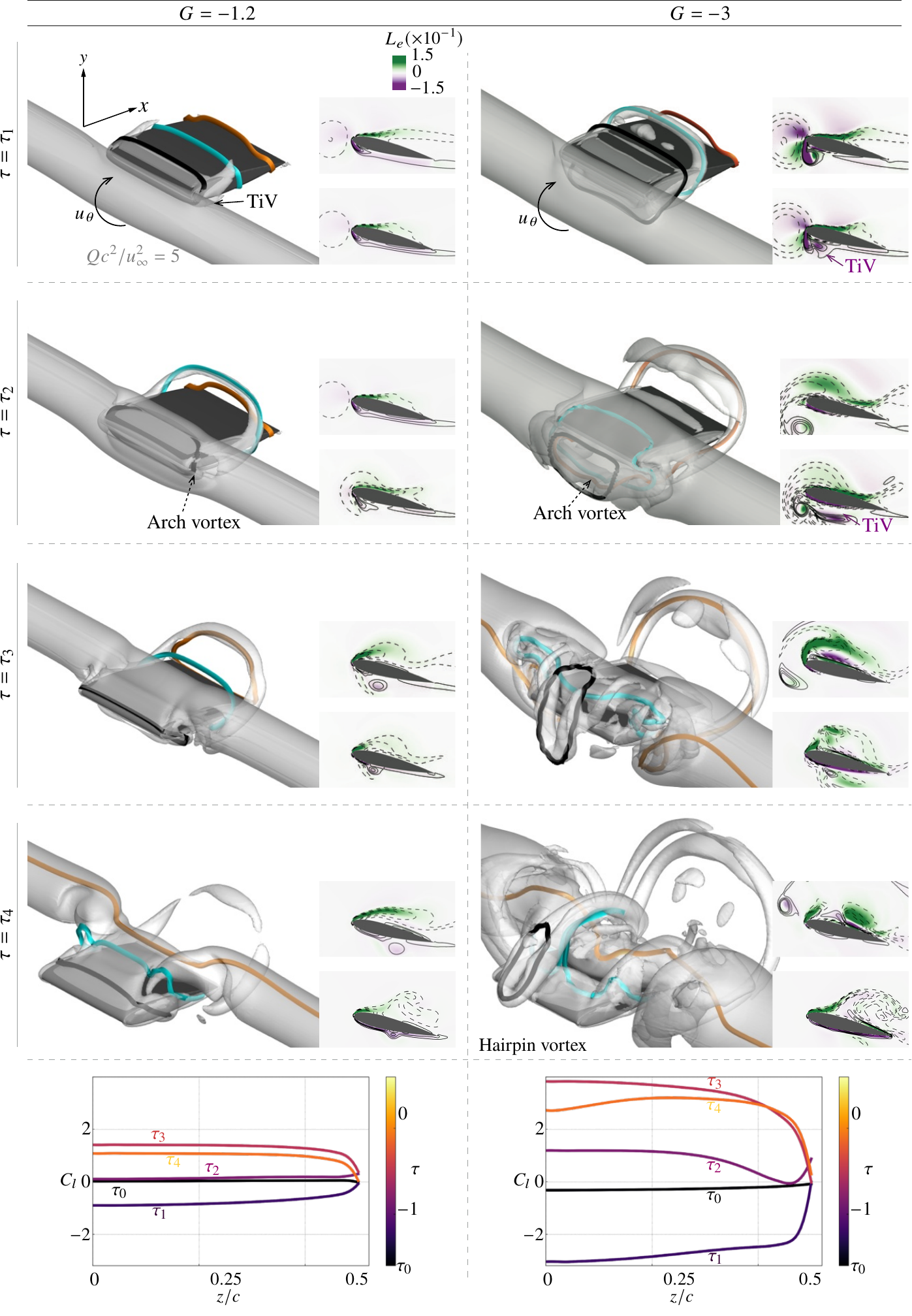}
   \caption{
   Same isosurface, spanwise slices, and sectional lift distributions as in figure~\ref{fig:fig4} for $G=-1.2$ and $-3$.}
\label{fig:fig5}
\end{center}
\end{figure}

\subsection{Negative vortex encounters by the square wing}
\label{subsec:3.3}
Let us analyse the negative gust vortex cases, shown in figure~\ref{fig:fig5}.
First, the evolution of the TiVs follows an opposite trend to that observed in the positive gust vortex cases. 
Specifically, TiVs first weaken and even reverse orientation, particularly near the leading edge, because the downward velocity of the approaching gust vortex applies higher pressure on the top surface and temporarily inverts the pressure and suction effects on the wing.
Note that the reversed TiVs induce local upwash effects, as can be seen in the sectional lift at~$\tau=\tau_1$ in the last row of figure~\ref{fig:fig5}, where the sectional lift exhibits a sharp increase near the wingtip.
An additional discussion on the upwash is presented in appendix~\ref{sec:appendix:a}.
Once the gust vortex passes, TiVs regain strength in their original orientations, affected by the upward velocity from the gust vortex.

This evolution is observed in the Q-criterion isosurface in figure~\ref{fig:fig5}.
At~$\tau=\tau_1$, the reversed TiVs around the leading edge are visible, while the TiVs near the trailing edge are too weak to be captured by the isosurface.
By~$\tau=\tau_4$, the TiVs recover in the original orientation, particularly near the leading edge.
The sectional lift plot in the last row of figure~\ref{fig:fig5} also reveals the effects of the TiV dynamics.
Due to the reversed TiVs, the sectional lift exhibits an increasing trend towards the tips at~$\tau=\tau_1$ for both $G=-1.2$ and $-3$.
Around~$\tau=\tau_3$, the sectional lift trend shifts back to a decreasing one, owing to a recovery of the TiVs in the original orientation.

We now analyse the three-dimensional interactions and connections between vortical structures.
Similar to the positive gust vortex cases but on the other side of the wing, the LEV and TiVs interconnect and form arch vortices below the wing, as seen at $\tau=\tau_2$ for both $G=-1.2$ and $-3$ in figure~\ref{fig:fig5}.
For $G=-1.2$, the vortices advect downstream along the wing over time, exhibiting kinks around the quarter spans on the upper surface at $\tau=\tau_4$ affected by the recovered TiVs.

For $G=-3$, the legs of the arch vortices in the inboard region are pushed toward and localized around the root by~$\tau=\tau_3$~(the vortex line in black), while the legs near the wingtips attach to the bottom surface of the wing~(the vortex line in aqua).
Furthermore, vortex lines connecting the TEV and reversed TiVs link to the gust vortex~(the vortex line in orange).
By~$\tau=\tau_4$, the arch vortex localized inboard is further pushed toward the root and stretched into a hairpin vortex.
The formation of this stretched hairpin vortex disrupts the redevelopment of the leading-edge shear layer on the upper surface, as seen in the spanwise slice at the root in figure~\ref{fig:fig5}.
After~$\tau=\tau_4$, the stretched hairpin vortex convects above the wing, although the subsequent evolution of the vortices is not shown in figure~\ref{fig:fig5}.

Next, let us identify the dominant vortical structures responsible for the large lift fluctuations through lift element analysis in figure~\ref{fig:fig5}.
Similar to the 2D wing cases, the LEV and shear layer beneath the wing primarily drives the first negative peak at~$\tau=\tau_1$, while the gust vortex merged with the leading-edge shear layer on the upper surface is the main contributor to the positive lift peak at~$\tau=\tau_3$.
For the square wing cases, the reversed TiVs also play a crucial role in negative lift generation by creating low-pressure cores below the wing. 
This effect is particularly visible for~$G=-3$ at~$\tau=\tau_1$ and~$\tau_2$, as shown in the lift element visualizations in figure~\ref{fig:fig5}, acting in the opposite manner to the strengthened TiVs in the positive vortex cases.
The negative spike in the section lift for~$G=-3$ at~$\tau=\tau_2$ in the last row of figure~\ref{fig:fig5} is also a consequence of the lift-decreasing effect by the low-pressure cores of the reversed TiVs.

Key structures that contribute to the attenuated lift response compared to the 2D wing can also be uncovered.
The first, negative peak is mitigated likely because of the upwash effects induced by the reversed TiVs.
For the attenuation of the positive peak, there appear to be two main mechanisms.
First, the arch vortex below the wing decreases lift by positioning the low-pressure regions closer to the bottom surface near the tips.
Second, the positive lift attributed to the gust vortex merged with the leading-edge shear layer on the top surface is reduced particularly near the wingtips.
This is because coherent spanwise vorticity of the gust vortex is partially lost as the gust vortex distorts around the wing corners, coupled with suppressed shear layer development near the tips, as shown in figure~\ref{fig:fig5}.

We thus find two opposing ways the finite wing affects the lift fluctuations for negative gust vortex cases:
(i)~the TiVs, which are low-pressure cores, develop below the wing through interaction with an impacting gust vortex, contributing to initial, large lift drop;
(ii)~the lift drop and surge are attenuated through the three-dimensional vortex dynamics such as the reversed TiVs and the distortion of the impacting gust vortex around the wing corners.
Reminiscent of the positive gust vortex cases, the second effect outweighs the first, causing the square wing to experience a smaller lift variation than the 2D wing.
These results suggest that further amplifying TiV evolution---intensifying TiVs in the opposite orientation to the baseline in response to a negative gust vortex---could enhance the attenuation of the large lift changes.

\subsection{Influence of gust vortex vertical position}
\label{subsec:3.4}

Let us study the effects of the relative height difference between the wing and the approaching gust vortex.
We first present lift fluctuations experienced by the 2D and square wings encountering a gust vortex introduced at different initial vertical positions.
Next, we examine the vortex dynamics around the 2D wing that contribute to the large lift changes for the varied vertical position cases.
We further analyse the flows around the square wing and investigate the finite-wing effects on the lift variations.

\begin{figure}
\begin{center}
   \includegraphics[width=1\textwidth, scale=1]{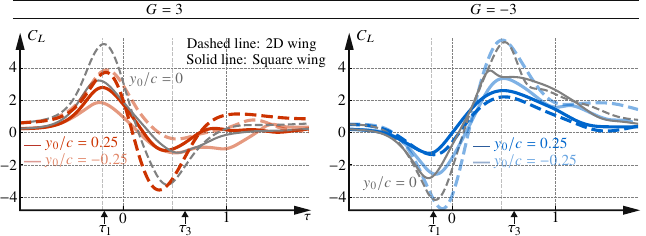}
   \caption{
   Lift history for the 2D~(dashed line) and square~(solid line) wings encountering a gust vortex with $G=3$~(left) and~$-3$~(right) initially introduced at different vertical positions~$y_0/c=\{-0.25,0,0.25\}$.
   }
\label{fig:fig6}
\end{center}
\end{figure}

Here, we compare the lift histories for the 2D and square wings subjected to a gust vortex with~$G=\pm3$ introduced at different initial vertical positions~$y_0/c=\{-0.25,0,0.25\}$, as shown in figure~\ref{fig:fig6}.
For the positive gust vortex, the 2D wing experiences the lowest lift fluctuation for the $y_0/c= -0.25$ case compared to the other two position cases, although the initial peak is slightly higher than that observed for the $y_0/c= 0.25$ case.
Similarly, the square wing exhibits the lowest lift fluctuation for the $y_0/c=-0.25$ case.
For example, the initial lift peak for the $y_0/c= -0.25$ case with the square wing is $1.67$ while the one for the $y_0/c= 0$ case is $2.70$.
For all three vertical positions considered herein, the lift fluctuations are attenuated for the square wing compared to those of the 2D wing.

For the negative gust vortex cases shown on the right of figure~\ref{fig:fig6}, lift fluctuations are the lowest when both the 2D and square wings fly beneath the gust vortex~($y_0/c=0.25$) among the three vertical position cases considered.
The 2D wing experiences the largest lift fluctuation when the vortex is located at $y_0/c= -0.25$, whereas the square wing exhibits the largest fluctuation at $y_0/c= 0$.
Notably, the lift peaks experienced by the square wing are reduced for all three examined vertical positions compared to the 2D wing.

\begin{figure}
\begin{center}
   \includegraphics[width=1\textwidth, scale=1]{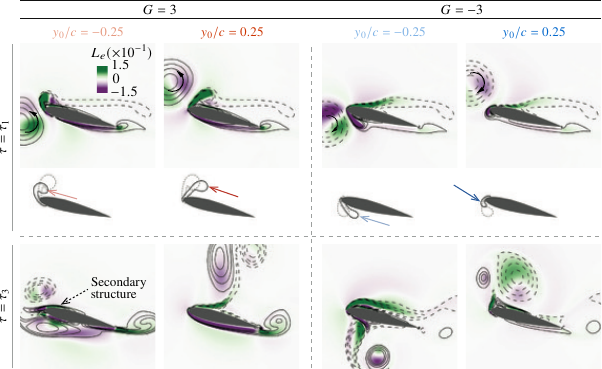}
   \caption{
   Lift element~$L_e$~(colour contour) and spanwise vorticity~$\omega_z$~(line contour) for the 2D wing cases of $G=\pm3$ with $y_0/c=\pm0.25$ at~$\tau=\tau_1$ and $\tau_3$.
   At~$\tau=\tau_1$, contour lines of $\omega_z=-20$~for $G=3$ and $\omega_z=20$~for $G=-3$ are also visualized below to compare the LEV between $y_0/c =0$~(dotted-line contour) and $\pm0.25$~(solid-line contour indicated by the coloured arrows).
   }
\label{fig:fig7}
\end{center}
\end{figure}

Let us shift our focus to the vortex dynamics around the 2D wing, particularly for~$y_0/c=\pm0.25$, and their relation to the lift fluctuations. 
In figure~\ref{fig:fig7}, we visualize lift element~$L_e$ and spanwise vorticity~$\omega_z$ for the 2D wing with~$y_0/c=\pm0.25$ at~$\tau=\tau_1$ and~$\tau_3$.
Below the visualization at~$\tau=\tau_1$, we also show the contour lines of $\omega_z=-20$ for $G=3$ and $\omega_z=20$ for $G=-3$ to compare the size and position of the LEV between the cases with $y_0/c=0$ and $\pm0.25$.

We examine the 2D wing flying above the positive gust vortex~$(G,y_0/c) = (3,-0.25)$, as shown in the first column of figure~\ref{fig:fig7}. 
As the gust vortex approaches the wing, a large LEV develops, similar to the positive gust vortex cases discussed in subsection~\ref{subsec:3.1}.
However, due to the lower vertical position of the gust vortex, the size and intensity of the LEV are weakened with its location shifted upstream compared to the $(G,y_0/c) = (3,0)$ case.
This weaker LEV results in a lower lift peak at~$\tau=\tau_1$ than the $y_0/c=0$ case.
In addition, at~$\tau=\tau_1$, a secondary vortical structure is induced between the LEV and the upper surface.
By~$\tau=\tau_3$, this secondary structure disrupts the vortical sheet feeding the LEV and rolls up around the leading edge.
Furthermore, the gust vortex convects below the wing and disturbs the bottom-surface boundary layer, resulting in a reduction of vorticity contained in it, as seen in the bottom left panel of figure~\ref{fig:fig7}.
This causes a loss of negative lift elements on the lower surface, substantially diminishing the negative lift peak, compared to the cases where the gust vortex is introduced higher; 
e.g., the negative lift peak for the $(G,y_0/c) = (3,-0.25)$ case of the 2D wing is $-0.49$ while the one for the $(G,y_0/c) = (3,0)$ case is $-3.10$.

Considering the 2D wing flying below the positive gust vortex $(G,y_0/c) = (3,0.25)$, as shown in the second column of figure~\ref{fig:fig7}, we observe similar vortex dynamics to the case with $(G,y_0/c) = (3,0)$ presented in subsection~\ref{subsec:3.1}.
That is, at~$\tau=\tau_1$, a large LEV forms on the upper surface while the approaching gust vortex increases the wall-normal velocity gradient in the bottom-surface boundary layer, enhancing vorticity generation.
However, compared to the $(G,y_0/c) = (3,0)$ case, the LEV has a smaller growth with its location shifted downstream, resulting in a lower positive lift peak.
By~$\tau=\tau_3$, the gust vortex convects downstream above the wing, forming a vortex-pair-like structure with the LEV.
As a result, the wing loses most of the vortical structures contributing to positive lift.
In contrast, vortical structures on the lower surface that contribute to negative lift persist, leading to a pronounced negative lift peak.

Let us next focus on the negative vortex encounter with $(G,y_0/c)=(-3,-0.25)$ by the 2D wing, as shown in the third column of figure~\ref{fig:fig7}.
As the negative vortex approaches the wing, a strong LEV develops below the wing, leading to a significant negative lift peak around~$\tau=\tau_1$.
Concurrently, a secondary vortical structure is induced between the LEV and the lower surface.
However, as the gust vortex convects below the wing by~$\tau=\tau_3$ forming a vortex-pair-like structure with the LEV, this secondary structure does not continue to grow, unlike the $(G,y_0/c)=(-3,0)$ case discussed in subsection~\ref{subsec:3.1}.
On the top surface, the gust vortex induces a large wall-normal velocity gradient in the shear layer near the leading edge by~$\tau=\tau_3$, generating a high level of vorticity, and hence, large positive lift.
This contributes to a significant second, positive lift peak.

For the 2D wing flying under the negative gust vortex~$(G,y_0/c)=(-3,0.25)$ as in the fourth column of figure~\ref{fig:fig7}, a smaller LEV develops below the wing at~$\tau=\tau_1$, compared to the cases where the wing flies higher. 
As a result, the first lift peak for this case is the smallest among the three examined different vertical position cases.
A secondary vortical structure induced between the LEV and the lower surface at~$\tau=\tau_1$ rolls up around the leading edge by~$\tau=\tau_3$, resembling the dynamics discussed for the $(G,y_0/c)=(-3,0)$ case in subsection~\ref{subsec:3.1}.
However, the positive lift elements associated with this secondary vortical structure around the leading edge are not as substantial as those observed in the $(G,y_0/c)=(-3,0)$ case because of the smaller development of the LEV, resulting in a lower positive lift peak than the $(G,y_0/c)=(-3,0)$ case.

Thus, we find effective vertical positions of the 2D wing to reduce large lift fluctuations against an impacting gust vortex;
flying over the positive gust vortex and flying under the negative gust vortex.
A similar trend is also observed in an experimental study by~\citet{peng2017asymmetric}.
They reported that the force fluctuation during vortex-gust encounters by a 2D wing with~$|G|<1$ at $Re=\mathcal{O}(10^5)$ in the context of blade-vortex interactions can be reduced when the airfoil passes above a positive vortex or below a negative vortex.

\begin{figure}
\begin{center}
   \includegraphics[width=1\textwidth, scale=1]{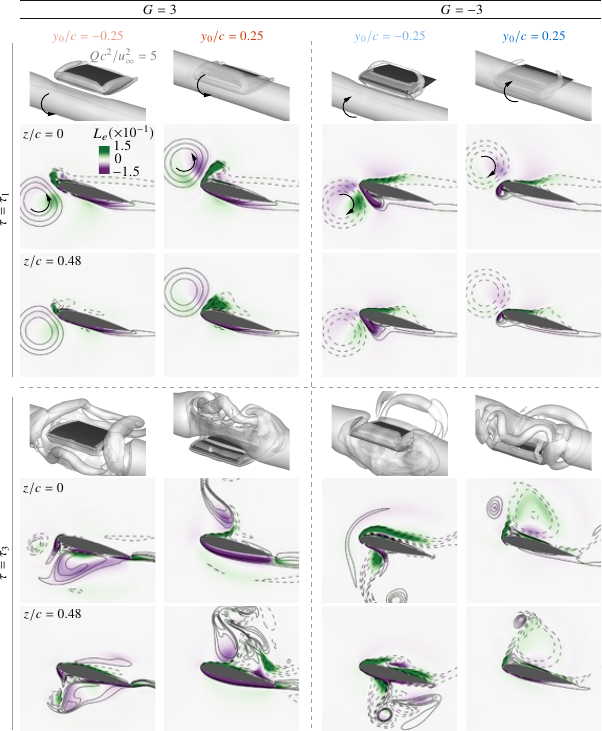}
   \caption{
Isosurfaces of Q criterion around the square wing and spanwise slices of $L_e$~(colour contour) and~$\omega_z$~(line contour) along the root and near the tip at~$\tau=\tau_1$ and $\tau_3$ for the $y_0/c=\pm0.25$ cases.
   }
\label{fig:fig8}
\end{center}
\end{figure}

We now turn our attention to the square wing cases.
Presented in figure~\ref{fig:fig8} are the isosurfaces of Q criterion and the spanwise slices of lift element~$L_e$ and spanwise vorticity~$\omega_z$ along the root and near the tip for the square wing with~$y_0/c=\pm 0.25$ at~$\tau=\tau_1$ and~$\tau_3$.

For~the $(G,y_0/c) = (3,-0.25)$ case, as shown in the first column of figure~\ref{fig:fig8}, a LEV develops on the upper surface at~$\tau=\tau_1$, connecting to the strengthened TiVs around the wing corners.
Similar to the discussion provided in section~\ref{subsec:3.2}, the strengthened TiVs suppress the development of the LEV and induce enhanced downwash, particularly near the wingtips, resulting in a smaller first lift peak than the 2D wing case.
By~$\tau=\tau_3$, the LEV transforms into an arch vortex, with its legs pushed toward and positioned beneath the wing near the tips.
The secondary vortical structure, which is induced between the LEV and the upper surface at~$\tau=\tau_1$, rolls up around the leading edge and shifts to the underside of the wing, particularly near the wingtip.
As in the 2D case, the gust vortex, convecting below the wing, interacts with the bottom-surface boundary layer, weakening negative lift elements on the lower surface.
Moreover, as seen in the spanwise slice at~$z/c=0.48$ at~$\tau=\tau_3$, the gust vortex loses some coherence in its spanwise structure near the tips, resulting in reduced negative-lift elements.

Let us next examine the square wing case with~$(G,y_0/c) = (3,0.25)$, as shown in the second column of figure~\ref{fig:fig8}.
As the gust vortex approaches the wing, a large LEV develops while the TiVs are amplified.
As seen in the spanwise slice of the lift element visualization along~$z/c=0.48$ at~$\tau=\tau_1$, the TiVs contribute to positive lift near the tip due to their large low-pressure cores, similar to the cases discussed in subsection~\ref{subsec:3.2}.
Concurrently, the intensified TiVs locally induce enhanced downwash and suppress the development of the LEV, resulting in a lower initial lift peak compared to the 2D wing case.
By~$\tau=\tau_3$, the gust vortex rolls up and lifts the arch vortex and TiVs significantly away from the wing, as visualized by the Q criterion isosurface.
Furthermore, spanwise vorticity generation in the bottom-surface boundary layer is suppressed near the tips, causing a loss of negative lift elements, similar to the discussion provided in subsection~\ref{subsec:3.2}.
These finite-wing effects contribute to attenuating the negative lift peak compared to the 2D wing.

For the square wing encountering the negative gust vortex with~$(G,y_0/c)=(-3,-0.25)$, as depicted in the third column of figure~\ref{fig:fig8}, a large LEV develops below the wing while the TiVs reverse orientation.
The reversed TiVs locally induce upwash effects and suppress the LEV development near the tips, attenuating the first negative lift peak compared to the 2D wing case.
Over time, the gust vortex rolls up the LEV and TiVs beneath the wing, disrupting the vortex sheet feeding the LEV, as visualized by the Q criterion isosurface and spanwise slices at~$\tau=\tau_3$.
On the upper surface, the development of the leading-edge shear layer, enhanced by the impacting gust vortex as discussed for the 2D wing case, weakens near the tips because the coherent structure of the gust vortex distorts around the wing corners.
The weaker leading-edge shear layer results in reduced positive lift elements, contributing to the damping of the second lift peak compared to the 2D wing case.

For the square wing case with~$(G,y_0/c)=(-3,0.25)$, as presented in the fourth column of figure~\ref{fig:fig8}, a smaller LEV forms below the wing at~$\tau=\tau_1$ than in the cases where the gust vortex is introduced lower.
The reversed TiVs develop beneath the wing, inducing upwash effects, suppressing the development of the LEV, and thereby mitigating the first lift peak compared to the 2D wing case.
By~$\tau=\tau_3$, the LEV forms an arch vortex, rolls up around the leading edge to the upper surface, and separates from the wing with the legs connected to the TiVs.
Similar to the 2D case, the secondary vortical structure, which is induced between the LEV and the bottom surface at~$\tau=\tau_1$, rolls up around the leading edge by~$\tau=\tau_3$.
As visualized in the spanwise slices at~$\tau=\tau_3$, the positive lift elements associated with this secondary vortical structure are reduced near the wingtips, because the development of the LEV was suppressed around~$\tau=\tau_1$ due to the reversed TiVs.
These attenuated positive lift elements contribute to a mitigated positive lift peak compared to the 2D wing case.

\begin{figure}
\begin{center}
   \includegraphics[width=0.8\textwidth, scale=1]{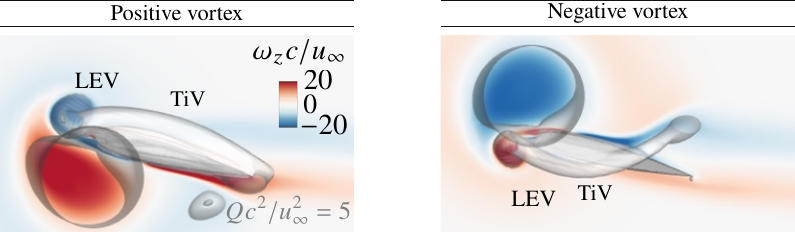}
   \caption{
    Side view of the Q criterion isosurface $Q=5$ and colour contours of the spanwise vorticity~$\omega_z$ along the root at~$\tau=0$ for the $(G,y_0/c)=(3,-0.25)$ and $(-3,0.25)$ cases with the square wing.
   }
\label{fig:fig9}
\end{center}
\end{figure}

Finally, to gain enhanced insights from the present results, we provide figure~\ref{fig:fig9}, where the side view of the Q criterion and spanwise vorticity~$\omega_z$ along the root at~$\tau=0$ are shown for $(G,y_0/c)=(3,-0.25)$ and $(-3,0.25)$ with the square wing.
For the case of the positive vortex encountering the square wing, as shown in figure~\ref{fig:fig9}~(left), a strong compact LEV develops on the top surface, leading to a significant positive lift peak.
Furthermore, the TiVs strengthen through interaction with the gust vortex and generate large low-pressure cores above the wing, contributing to substantial vortex lift near the wingtips.
However, the strengthened TiVs locally induce enhanced downwash and suppress further LEV development, attenuating the large lift peak compared to the 2D wing.
Notably, the large lift fluctuation is reduced when the square wing flies over the positive gust vortex as visualized in figure~\ref{fig:fig9}~(left), due to smaller LEV development and a loss of negative lift elements in the bottom-surface boundary layer by the impact.

Conversely, when the negative gust vortex encounters the square wing, as presented in figure~\ref{fig:fig9}~(right), a strong compact LEV and reversed TiVs develop below the wing, resulting in a significant negative lift peak.
Simultaneously, the reversed TiVs locally induce upwash and suppress LEV growth, attenuating the lift peak compared to the 2D wing.
Opposite to the aforementioned positive gust vortex case, the large lift fluctuation caused by the negative gust vortex can be reduced by the wing flying under the gust vortex, as visualized in figure~\ref{fig:fig9}~(right).
These findings inform potential strategies for lift attenuation during extreme gust-finite-wing interactions by taking advantage of wing positions and TiVs.

\section{Concluding remarks}
\label{sec:conclusion}

This study identified key vortex dynamics that generate large lift fluctuations in extreme vortex-gust encounters by a square-planform wing and cause lift attenuation when compared to the case of a 2D wing.
We conducted direct numerical simulations of flows over an $sAR=0.5$ wing at $Re=600$ and performed lift element analysis on the unsteady flowfield.
We uncovered that TiVs play two opposing roles in the large lift changes during extreme vortex gust encounters by the finite wing;
1)~contributing to the lift surge or drop by developing large low-pressure cores near the wingtips;
2)~mitigating the lift variations through three-dimensional vortex dynamics, such as their enhanced downwash or upwash.
The second effect dominates the first, causing the square wing to undergo a smaller lift change than the 2D wing.
In addition, we revealed effective vertical positions of the wing to reduce the lift change against an incoming gust vortex---flying over the positive gust vortex and flying beneath the negative one.
The current insights obtained in the laminar flow setting can establish a foundation for extreme gust-finite-wing interacting flows and direct future efforts in developing control strategies or wing configurations to leverage TiV dynamics and mitigate large lift fluctuations at higher Reynolds numbers.

\appendix

\section{Enhanced downwash and upwash of the TiVs}
\label{sec:appendix:a}

\begin{figure}
\begin{center}
   \includegraphics[width=1\textwidth, scale=1]{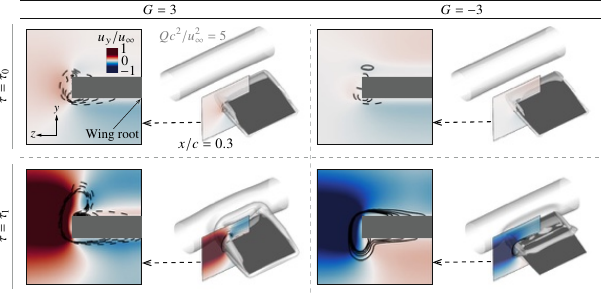}
   \caption{
   Transverse velocity~$u_y$~(color contours) and streamwise vorticity~$\omega_x$~(line contours) along the streamwise slice~$x/c=0.3$ at~$\tau=\tau_0$ and~$\tau_1$ for the cases of $G=3$ and~$-3$ with~$y_0/c=0$.
   Solid-line contours correspond to positive streamwise vorticity while dashed-line contours correspond to negative streamwise vorticity.
   }
\label{fig:fig10}
\end{center}
\end{figure}

We present in figure~\ref{fig:fig10} the transverse velocity and streamwise vorticity along the streamwise slice~$x/c=0.3$ at~$\tau=\tau_0$ and~$\tau_1$ for the cases of $G=\pm3$ with~$y_0/c=0$.
For $G=3$, the TiV grows in size and intensity from~$\tau=\tau_0$ to~$\tau_1$.
Furthermore, we observe a stronger downward velocity above the wing at~$\tau=\tau_1$ than~$\tau_0$, indicating that the downwash of the TiVs are locally enhanced.

Regarding~the $G=-3$ case, we can see coherent positive streamwise vorticity below the wing near the tip---the reversed TiV---at~$\tau=\tau_1$, which are not present at $\tau=\tau_0$.
Consequently, the transverse velocity below the wing switches from negative at~$\tau=\tau_0$ to positive at~$\tau=\tau_1$, suggesting an upwash effect of the reversed TiVs.

\section{Verification and validation}
\label{sec:appendix:b}

\begin{figure}
\begin{center}
   \includegraphics[width=1\textwidth, scale=1]{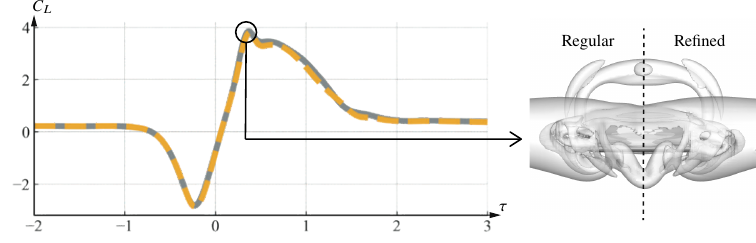}
   \caption{
   Lift coefficient change for the $(G,y_0/c) = (-3,0)$ case with the regular and refined meshes.
   A front view of the instantaneous Q criterion isosurface~$Qc^2/u_\infty^2=6$ around the second lift peak is also shown for the two different meshes.
   }
\label{fig:fig11}
\end{center}
\end{figure}

We verify the grid convergence of the current mesh against a refined mesh.
The current mesh, referred to as a regular mesh, consists of approximately $4.9\times 10^6$ control volumes.
The refined mesh comprised of approximately $2.4 \times 10^7$ control volumes includes increased resolution in all three spatial directions around the wing.
Grid convergence is assessed by comparing the lift coefficient change and the instantaneous Q-criterion isosurface during the gust encounter with $(G,y_0/c) = (-3,0)$, as shown in figure~\ref{fig:fig11}.
We also validate the current mesh and numerical simulation with the time-averaged lift coefficient examined in~\citet{zhang_rapid} and~\citet{ribeiro2023laminar}, as shown in table~\ref{tab:validation}.
Based on these agreements, the regular mesh is deemed sufficient to perform the direct numerical simulations in this study.

\begin{table}
\begin{center}
\begin{tabular}{ccccccc}
                     &                      &                      &                      &                      Reference            & Present              \\
                     & $sAR$                & $\alpha$             & $Re$                               & $\bar{C_L}$          & $\bar{C_L}$          \\ \hline
\citet{zhang_rapid}                 &         0.5             &         $20^\circ$             &           400                           &     0.32                 &          0.31           \\
\citet{ribeiro2023laminar}                &             2         &           $14^\circ$           &         600                         &        0.35              &       0.35               \\
\multicolumn{1}{l}{} & \multicolumn{1}{l}{} & \multicolumn{1}{l}{} & \multicolumn{1}{l}{} & \multicolumn{1}{l}{} & \multicolumn{1}{l}{} & \multicolumn{1}{l}{}
\end{tabular}
\caption{
Validation of the present simulation.
\citet{zhang_rapid} use an incompressible flow solver whereas the current study uses the compressible flow solver.
}
\label{tab:validation}
\end{center}
\end{table}

\section*{Acknowledgments}
This work was supported by the US Department of Defense Vannevar Bush Faculty Fellowship (N00014-22-1-
2798) and the Air Force Office of Scientific Research (FA9550-21-1-0174). 
H.O.acknowledges partial support from
Honjo International Scholarship Foundation.

\bibliographystyle{plainnat}
\bibliography{references}  

\end{document}